\documentclass{jaa}
\usepackage{graphicx}

\begin{document}\sloppy
\title{Abundances of neutron-capture elements in CH and Carbon-Enhanced Metal-Poor (CEMP) stars}
\author{Meenakshi Purandardas\textsuperscript{1} and Aruna Goswami\textsuperscript{1}}
\affilOne{\textsuperscript{1}Indian Institute of Astrophysics, Koramangala, Bangalore 560034, India\\}
\twocolumn[{
\maketitle
\corres{meenakshi.p@iiap.res.in, aruna@iiap.res.in}
\msinfo{  ---- March  2020}{--------- }

\begin{abstract}
All the elements heavier than Fe are produced either by slow (-s) or 
rapid (-r) neutron-capture process. 
Neutron density prevailing in the stellar sites is one of the major factors 
that  determines the type of neutron-capture processes. We  
present the results based on the estimates of corrected value of absolute carbon abundance,
[C/N] ratio, carbon isotopic ratio and [hs/ls] ratio obtained from the 
high resolution spectral analysis of six  stars that 
include both  CH stars and CEMP stars. All  the stars show enhancement of
neutron-capture elements. Location of these objects in the 
A(C) vs. [Fe/H] diagram shows that they are Group I objects, with external 
origin of carbon and neutron-capture elements. Low values of carbon isotopic 
ratios estimated for  these objects may also be  attributed to some 
external sources.  As the carbon isotopic ratio is a good indicator of mixing, 
we have used the estimates of  $^{12}$C/$^{13}$C ratios  to examine the occurance of  mixing 
in the stars. While the object HD 30443 might have  experienced an 
extra mixing process that usually occurs after 
red giant branch (RGB) bump for stars with log(L/L$_{\odot}$) $>$ 2.0, the  
remaining objects do not show any  evidence of  having undergone 
any such  mixing process.
The higher values of [C/N] ratios obtained for these objects also indicate that 
none of these objects have experienced any strong internal mixing processes. 
Based on the estimated abundances of carbon  and  the neutron-capture
elements, and the  abundance ratios, we have classified   the objects  into 
different groups. While the objects HE 0110$-$0406, HD 30443 and CD$-$38 2151
are found to be CEMP-s stars, HE 0308$-$1612 and HD 176021 show characteristic
properties of CH stars with moderate enhancement of carbon. The object 
CD$-$28 1082 with enhancement of both r- and s-process elements is found 
to belong to the  CEMP-r/s group.
\end{abstract}


}]
\doinum{12.3456/s78910-011-012-3}
\artcitid{\#\#\#\#}
\volnum{000}
\year{0000}
\pgrange{1--}
\setcounter{page}{1}
\lp{1}

\section{Introduction}
Chemical analysis of metal-poor stars such as CH stars and CEMP stars can provide important clues about the nature of nucleosynthesis processes occured in the early Galaxy. Especially, the abundances of neutron-capture elements can be used to constrain the Galactic chemical evolution due to heavy elements. Various sky survey programmes (HK survey and Hamburg/ESO survey, Beers {\em et al.} 1985; Wisotzki {\em et al.} 2000; Christlieb {\em et al.} 2001) were conducted in the past to find metal-poor stars. All these surveys show that the fraction of carbon enhanced objects increases with decreasing metallicity (Beers \& Christlieb 2005; Frebel {\em et al.} 2005; Norris {\em et al.} 2007; Spite {\em et al.} 2013; Yong {\em et al.} 2013). 

\par CH stars are FGK giants that show strong carbon molecular bands in their spectra. They are high radial velocity objects, mostly found in the halo of our Galaxy. CEMP stars are the metal-poor ([Fe/H] $<$ -1) counter parts of CH stars. Both the CH stars and CEMP stars show enhancement of carbon and neutron-capture elements. Hence these objects are ideal candidates to study the origin and evolution of these elements. Based on the type of enhancement of neutron-capture elements CEMP stars are classified into different groups, such as CEMP-s, CEMP-r, CEMP-r/s and CEMP-no stars (Beers \& Christlieb 2005). The evolutionary status of CH and CEMP stars do not support the enhancement of carbon and heavy elements observed in these stars. The widely accepted scenario to explain this enhancement is that these objects are in a binary system. The primary companion once passed through the Asymptotic Giant branch (AGB) phase and synthesized carbon and heavy elements. The synthesized materials are then transferred to the secondary companion through some mass transfer mechanisms. The radial velocity variations exhibited by CH and CEMP stars (McClure 1983, 1984;  McClure \& Woodsworth 1990; Hansen {\em et al.}2016a) support this idea.

\par In this paper, we have presented the results from the high-resolution analysis of six stars that include four CEMP stars and two CH stars. The paper is organized as follows. In section 2, we have presented a brief discussion on the new results obtained for our programme stars as some of the results from abundance analysis of these objects were presented in Purandardas {\em et al.} (2019), and Purandardas, Goswami \& Doddamani (2019). Section 3 presents the sample selection, observations and data reductions. Section 4 describes the determination of radial velocity and stellar atmospheric parameters. Details of the abundance anaysis are presented in section 5. In section 6, interpretation of our results are presented. Conclusions are drawn in Section 7.
\section{Novelty of this work}
We have presented the abundance analysis results for 26 elements in our programme stars in Purandardas {\em et al.} (2019), and Purandardas, Goswami \& Doddamani (2019). In these works, we have also reported the mass and age of these stars as well as the results from the kinematic analysis. The location of these objects in the H-R diagram shows that they are either sub-giants or in the ascending stage of the giant branch. Various mixing processes have been found to operate in giant stars. It is therefore important to understand whether the stars have undergone any internal mixing processes before interpretting the observed abundances. We had not addressed this problem in our previous works. In this paper, we have checked whether any internal mixing processes have altered the surface chemical composition of these stars based on [C/N] and carbon isotopic ratios. While  HD 30443 might have experienced an extra mixing process that usually occurs after red giant branch (RGB) bump for stars with log(L/L$_{\odot}$) $>$ 2.0, the remaining objects do not show any evidence of having undergone any such mixing process. In our previous works, we have not checked the possible source of enrichment of neutron-capture elements observed in our programme stars. In the present work, we have tried to understand the possible source of neutron-capture elements based on the location of the absolute carbon abundance values of the programme stars, in the A(C) vs. [Fe/H] diagram. Based on the estimated value of carbon abundances together with the observed  enhancement of neutron-capture elements, we have  classified them as Group I objects following Yoon {\em et al.} (2016) classification scheme. As the Group I objects    are all binaries, it is  likely  that our programme stars that belong to this group are also in binary systems with external origin of carbon and neutron-capture elements. Low values of carbon isotopic ratios estimated for these objects may also be attributed to external sources. In the present work, we have re-calculated the [hs/ls] ratio for our programme stars without considering the contribution from samarium which is an r-process element which we had taken into account for this estimation in our previous works.

\section{Observations and Data reduction}
Programme stars are selected from the CH star catalogue of Bartkevicious (1996), Goswami (2005) and Goswami {\em et al.} (2010). In the later papers, potential CH star candidates are identified based on the low resolution 
(R=$\lambda/\delta\lambda$$\sim$1330) spectroscopic studies of faint high latitude carbon stars. In these works, two of our programme stars HE 0110$-$0406 and 
HE 0308$-$1612 are identified  to be  potential CH star candidates. We have 
obtained  high resolution (R $\sim$ 60 000) spectra of  these objects along 
with HD 30443 using the high-resolution fiber fed Hanle Echelle Spectrograph 
(HESP) attached to the 2 m Himalayam Chandra Telescope (HCT) at the Indian 
Astronomical Observatory, Hanle. The spectra cover the wavelength range from 
3530 to 9970 {\rm \AA}. The spectrograph allows a resolution of 60 000 with 
slicer, and a resolution of 30 000 without slicer. The spectrum is recorded on 
a CCD with 4096$\times$4096 pixels of 15 micron size. For the programme stars, 
CD$-28$ 1082 and HD 176021, we have used the high resolution FEROS spectra (Fiber-fed Extended Range Optical Spectrograph (FEROS) of 1.52 m telescope of 
Europian Southern Observatory at La Silla). The wavelength coverage of the 
FEROS spectra is from 3500 - 9000 {\rm \AA} with a spectral resolution of 
$\sim$ 48 000. The detector is a back-illuminated CCD with 2948$\times$4096 
pixels of 15 $\mu$m size. For the object CD$-38$ 2151, a high resolution  
(R $\sim$ 72 000) spectrum was obtained using the  high-resolution fiber 
fed Echelle spectrometer attached to the 2.34 m Vainu Bappu Telescope (VBT) 
at the Vainu Bappu Observatory (VBO), Kavalur. The spectrum  covers the 
wavelegth region from 4100 to 9350 {\rm \AA} with gaps between orders. 
The spectrometer operates in two modes.  It allows a resolution of 72 000 
with a 60 micron slit and a resolution of 27 000 without the slit. The spectrum
is recorded on a CCD with 4096$\times$4096 pixels of 12 $\mu$m size. The data 
reduction is carried out using various spectroscopic reduction packages such 
as IRAF. Examples of a few sample spectra of the programme stars are shown 
in Figure 1.
\begin{figure}[htb]
\centering
\includegraphics[width=8.5cm,height=12cm]{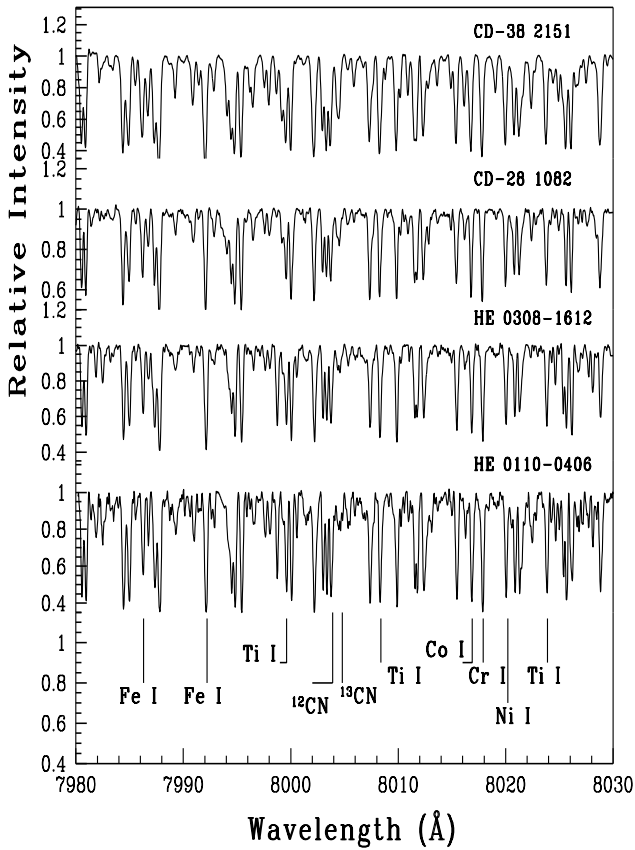}
\caption{ Sample spectra of the programme stars in the wavelength region 7980-8030 {\rm {\rm \AA}}. }
\end{figure}

\section{Determination of radial velocity and stellar atmospheric parameters}
Radial velocity of the programme stars are determined by measuring the shift in the wavelength for a large number of unblended and clean lines in their spectra. The radial velocities range from $-26.7$ to 139.7 km s$^{-1}$. The estimated radial velocities of the programme stars are presented in Table 1.  

\par Stellar atmospheric parameters are determined from the measured equivalent widths of clean and unblended Fe I and Fe II lines using the local thermodynamic equilibrium (LTE) analysis. We made use of the recent version of MOOG of Sneden (1973) for our analysis. Model atmospheres are selected from Kurucz grid of model atmospheres with no convective overshooting (http://cfaku5.cfa.hardvard.edu/). Solar abundances are taken from Asplund, Grevesse \& Sauval (2009). Effective temperature is taken to be that value for which the trend between the abundance derived from Fe I lines and the corresponding excitation potential gives a zero slope. At this temperature, microturbulent velocity is determined for which the abundance derived from Fe I lines do not exhibit any dependence on the reduced equivalent width. Corresponding to these values of effective temperature and microturbulent velocity, log g is determined in such a way that the abundances obtained from Fe I and Fe II lines are nearly the same. Only those lines with excitation potential from 0.0 - 5.0 eV and equivalent widths from 20 - 180 m{\rm \AA} are considered for the analysis. The derived atmospheric parameters and the radial velocities are listed in Table 1.
\vspace{-2em} 
\begin{table}
\caption{\bf Derived atmospheric parameters and radial velocities of the programme stars. }
\resizebox{\columnwidth}{!}{\begin{tabular}{lccccccc}
\topline
Star         &T$_{eff}$& log g &$\zeta$      & [Fe I/H]        &[Fe II/H]        & V$_{r}$       \\
             &    (K)  & (cgs)  &(km s$^{-1}$) &                 &                 & (km s$^{-1}$) \\
\hline
HE 0110$-$0406 & 4670 & 1.00 & 1.92          & $-1.31$$\pm$0.09& $-1.29$$\pm$0.12 & $-44.40$$\pm$3.8 (HESP) \\ 
HE 0308$-$1612& 4600   &1.70  &1.42          &$-0.72$$\pm$0.19 &$-0.73$$\pm$0.15 & 85.5$\pm$1.22 (HESP)\\
CD$-28$ 1082 & 5200    &1.90  &1.42          &$-2.46$$\pm$0.08 &$-2.44$$\pm$0.02 &  $-26.7$$\pm$0.3 (FEROS)\\
HD 30443     & 4040    & 2.05 &2.70          &$-1.68$$\pm$0.05 &$-1.69$$\pm$0.11 & 66.61$\pm$0.20 (HESP)\\
CD$-38$ 2151 & 4600    &0.90  &2.30          &$-2.03$$\pm$0.10 &$-2.03$          & 139.7$\pm$1.9 (VBT)\\
HD 176021    & 5900    &3.95  &1.02          &$-0.62$$\pm$0.08 &$-0.65$$\pm$0.05 & 109.1$\pm$0.5 (FEROS)\\ 
\hline
\end{tabular}}
\end{table}

\section{Abundance analysis}
The detailed discussion on the abundance analysis results for the six programme stars are presented in Purandardas {\em et al.} (2019), and Purandardas, Goswami \& Doddamani (2019). Here we present a brief summary of these results. However here we give more emphasize to the results based on the absolute carbon abundance, [C/N] ratio, carbon isotopic ratio and the [hs/ls] ratio which is recalculated in this work. 

\par The abundances of various elements are determined from the measured equivalent widths of absorption lines due to neutral and ionized elements. We have used only the symmetric and clean lines for our analysis. Lines are identified by overplotting the arcturus spectra upon the individual spectra of our programme stars. Then a master linelist is prepared using the measured equivalent widths and other line information such as lower excitation potential and the loggf values taken from the Kurucz database. We have also consulted VALD database. We could estimate the abundances of 24 elements which include the light elements C, N, O, odd-Z element Na, $\alpha$ - and Fe-peak elements Mg, Si, Ca, Ti, V, Cr, Mn, Co, Ni and Zn and the neutron-capture elements Sr, Y, Zr, Ba, La, Ce, Pr, Nd, Sm and Eu. We have also used spectrum synthesis calculations for elements such as Sc, V, Mn, Ba, La and Eu taking their hyperfine structures into considerations. The hyperfine structures of Sc, V and Mn are taken from Prochaska \& McWilliam (2000). For Ba, La and Eu, the hyperfine structures are taken from McWilliam (1998), Jonsell {\em et al.} (2006) and Woorley {\em et al.} (2013) respectively.

\par We could estimate oxygen abundance only for CD$-38$ 2151 and HD 30443. For these objects, the abundance of oxygen is determined from the spectrum synthesis calculations of the [OI] line at 6300.3 and OI line at 6363.8 {\rm \AA}.  The carbon abundance could be determined for all the objects from the spectrum synthesis calculations of the C$_{2}$ molecular band at 5165 {\rm \AA}.  We could estimate the carbon isotopic ratio for all of our programme stars except HD 176021 using the spectrum synthesis calculation of CN band at 8005 {\rm \AA} (Figure 2). The values lie in the range from 7.4 to 45. Abundance of nitrogen is estimated using the spectrum synthesis calculation of CN band at 4215 {\rm \AA}. Nitrogen is found to be enhanced in CD$-28$ 1082 and CD$-38$ 2151. Other objects exhibit moderate enhancement in nitrogen. The molecular lines for C$_{2}$ and CN are taken from Brooke {\em et al.} (2013), Sneden {\em et al.} (2014) and Ram {\em et al.} (2014). The estimated values of carbon and nitrogen are presented in Table 2.

{\footnotesize
\begin{table}[htb]
\caption{\bf Abundance results for carbon, nitrogen, C/O and carbon isotopic ratios. }
\resizebox{\columnwidth}{!}{\begin{tabular}{lccccccc}
\hline
Star           & log$\epsilon$(C) & log$\epsilon$(C)$^{\ast}$ & [C/Fe] & log$\epsilon$(N) & [N/Fe] & C/O  & $^{12}$C/$^{13}$C   \\

\hline
HE 0110$-$0406 & 7.85 & 7.98 &0.73 & 7.15 & 0.63 &-    & 45.0    \\
HE 0308$-$1612 & 8.48 & 8.50 &0.78 & 7.25 & 0.15 &-    & 15.6  \\
CD$-28$ 1082   & 8.16 & 8.23 &2.19 & 8.10 & 2.73 &-    & 16.0  \\
HD 30443       & 8.43 & 8.47 &1.68 & 6.55 & 0.40 &1.02 & 7.40  \\
CD$-38$ 2151   & 7.90 & 8.02 &1.50 & 7.20 & 1.40 &2.95& 11.2  \\
HD 176021      & 8.33 & 8.33 &0.52 & 7.80 & 0.59 &-   & -    \\
\hline
\end{tabular}}
{\tiny $^{\ast}$ Corrected value of carbon.}
\end{table}
}

\begin{figure}[htb]
\centering
\includegraphics[width=6.8cm,height=6cm]{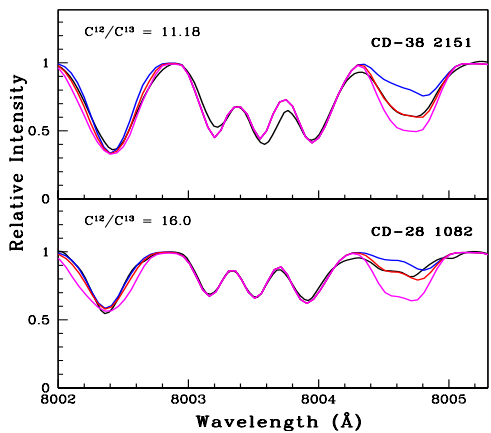}
\caption{ Synthesis of CN band around 8005 {\rm {\rm \AA}}. Synthesized spectra is shown in red colour and the observed spectra is represented in black colour. Synthetic spectra corresponding to $^{12}$C/$^{13}$C $\simeq$ 12 (blue) and 1 (magenta) are also shown. }
\end{figure}

\par Sodium is moderately enhanced in all our programme stars except HD 176021 in which Na is near solar. HE 0110$-$0406 and HD 176021 show near solar abundances of alpha elements. While, HE 0308$-$1612 shows slight enhancement of these elements. We could not estimate Si in these objects. Among the alpha elements, we could estimate only Mg and Ca in CD$-28$ 1082 with [Mg/Fe] $\sim$ 0.45 and [Ca/Fe] $\sim$ 0.27. In HD 30443, Si and Ca are moderately enhanced with [Si/Fe] $\sim$ 0.82 and [Ca/Fe] $\sim$ 0.51. While Magnesium, Sc and Ti are near solar. Magnesium and Ca are found to be moderately enhanced in CD$-38$ 2151. While Si is found to be enhanced with [Si/Fe] $\sim$ 1.62. Scandium and Ti are found to be near solar in this object. 

\par HE 0110$-$0406 shows near solar abundance of Fe peak elements except Ni which is slightly enhanced with [Ni/Fe] $\sim$ 0.45. While Fe-peak elements are slightly enhanced in HE 0308$-$1612. We could not estimate Co in HE 0308$-$1612. Among the Fe peak elements, we could estimate only Mn in CD$-28$ 1082 which is found to be enhanced with [Mn/Fe] $\sim$ 1.48. In HD 30443, Mn and Co are near solar and Ni is moderately enhanced with [Ni/Fe]$\sim$ 0.59. Cobalt and Ni are near solar in CD$-38$ 2151. While chromium is slightly enhanced and Mn is underabundant with [Mn/Fe] $\sim$ $-0.20$. In HD 176021, all the Fe peak elements are found to be near solar.

\par  All of our programme stars exhibit enhancement of neutron-capture 
elements. From our detailed analysis, we found that CD$-28$ 1082 is a 
CEMP-r/s star and the objects HE 0110$-$0406, CD$-38$ 2151 and HD 30443 
are CEMP-s stars. We could not estimate Eu in HD 30443 and CD$-38$ 2151.
Hence, it is not possible to classify these stars based on the criteria 
as given by Beers \& Christlieb (2005). In this case, we have used the 
criteria for the classification of CEMP-s stars as given by 
Hansen {\em et al.} (2019) based on [Sr/Ba] ratio. According to this 
classification scheme, [Sr/Ba] $>$ $-0.5$ can be used to seperate CEMP-s 
stars from CEMP-r/s stars. HD 30443 and CD$-38$ 2151 show [Sr/Ba] $\sim$ $-0.27$
and [Sr/Ba] $\sim$ 0.88 respectively. HE 0308$-$1612 and HD 176021 exhibit 
the properties of CH subgiants. All the programme stars show the enhancement 
of heavy s-process elements more than the light s-process elements except 
for CD$-38$ 2151 and HD 176021. The abundance results for neutron-capture 
elements are listed in Table 3. In this table, ls stands for light s-process 
elements (Sr, Y and Zr) and hs represents the heavy s-process elements 
(Ba, La, Ce, and Nd).

{\footnotesize
\begin{table}[htb]
\caption{\bf Ratios of light and heavy s-process elements }
\resizebox{\columnwidth}{!}{\begin{tabular}{lcccc}
\hline
Star         & [Fe/H] & [ls/Fe] & [hs/Fe]  & [hs/ls] \\

\hline
HE 0110$-$0406 & $-1.30$& 1.03 & 1.36 & 0.33 \\
HE 0308$-$1612 & $-0.73$& 1.11 & 1.62 & 0.51 \\
CD$-28$ 1082   & $-2.45$& 1.52 & 1.90 & 0.38  \\
HD 30443       & $-1.69$& 1.24 & 1.93 & 0.69  \\
CD$-38$ 2151   & $-2.03$& 1.24 & 1.10 &$-0.14$ \\
HD 176021      & $-0.64$& 1.50 & 1.37 &$-0.13$ \\
\hline
\end{tabular}}
\end{table}
}

\section{Interpretation of results}
The interpretation of the results based on the corrected value of absolute carbon abundance, 
[C/N], carbon isotopic ratios and [hs/ls] ratios obtained for our programme stars are presented here in detail.
Understanding the possible source of origin of the enhancement of neutron-capture
elements is very important to understand the type of nucleosynthesis process which produced it.
One of the best ways to get any clues regarding the source is to locate the stars in the
A(C) vs. [Fe/H] diagram. It is found  that CEMP stars exhibit a bimodal distribution in the 
A(C) vs. [Fe/H] diagram. Spite {\em et al.} (2013) claims the occurance of 
two plateau at A(C) = 8.25 and another at A(C) = 6.50. While Yoon {\em et al.} 
(2016) confirmed two peaks at A(C) = 7.96 and A(C) = 6.28 respectively for 
high and low carbon regions corresponding to the corrected carbon values. 
The stars that occupy the high carbon region are mostly found to be in binary 
systems. Hence the observed enhancement of carbon is attributed to the binary 
companion. Stars that occupy the low carbon region are found to be single 
and the observed carbon abundance is intrinsic in origin. CEMP stars are 
classified by Yoon {\em et al.} (2016) into three groups based on the 
morphology in the A(C) vs. [Fe/H] diagram. Group I objects are mainly 
composed of CEMP-s and CEMP-r/s stars. These objects show a weak dependence 
of A(C) on [Fe/H]. The absolute carbon abundance, A(C) of Group II objects 
show a clear dependence on [Fe/H]. While for Group III objects, A(C) is 
found to be independent of [Fe/H], Group II and Group III objects are mainly 
composed of CEMP-no stars. In Figure 3, the location of our programme stars 
are shown in the A(C) vs. [Fe/H] diagram. We have applied corrections to 
the estimated carbon abundances using the public online tool by 
Placco  {\em et al.} (2014) available at http/://vplacco.pythonanywhere.com/. 
The corrected carbon values are listed in Table 2. Figure 3 shows that all 
of our programme stars are Group I objects. Hence we assume that the 
observed enhancement of carbon may be attributed to the binary companion. 

\begin{figure}[htb]
\centering
\includegraphics[width=8cm,height=8cm]{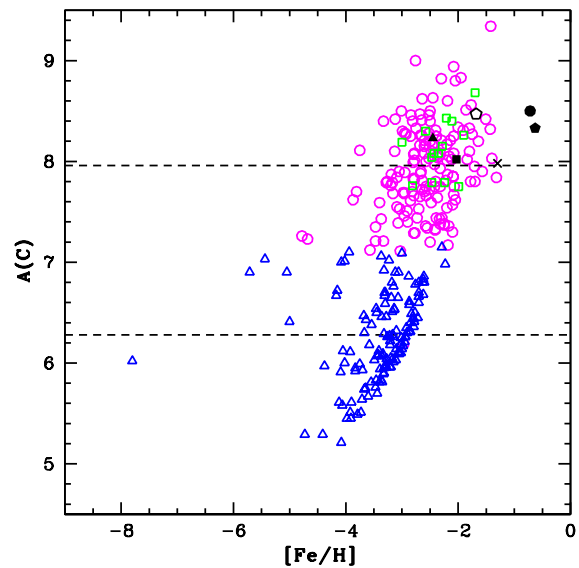}
\caption{Corrected A(C) vs. [Fe/H] diagram for the compilation of CEMP stars taken from Yoon {\em et al.} (2016). CEMP-s stars are represented by Open circles. CEMP-r/s and CEMP-no stars are shown by open squares and open triangles respectively. Programme stars are represented by black coloured symbols, HE 0110$-$0406 (cross), HE 0308$-$1612 (filled circle),  CD$-28$ 1082 (filled triangle), HD 30443 (open pentagon), CD$-38$ 2151 (filled square) and HD 17602 (filled pentagon)}
\end{figure}

\par The low values of $^{12}$C/$^{13}$C ratio observed in our programme 
stars also support the extrinsic origin of carbon. Various mixing processes 
such as first dredge up, thermohaline mixing and rotation induced mixing 
(Charbonnel 2005; Dearborn {\em et al.} 2006; Eggleton {\em et al.} 2006) 
can cause a low carbon isotopic ratio. Vanture (1992) suggests that certain 
nucleosynthetic reactions of the accreted material  can 
reduce the $^{12}$C abundance. We have estimated the [C/N] ratio of our 
programme stars to derive any clues  if  they have undergone any mixing 
processes. Since nitrogen is produced at the expense of carbon, [C/N] 
ratio acts as a sensitive indicator of mixing. Spite {\em et al.} (2005) 
analysed CNO and Li abundances for a sample of extremely metal-poor stars. 
Their analysis shows that the stars that exhibit clear evidence of mixing 
have a low value of [C/N] ratio ($<$ $-0.60$), and they belong to upper RGB 
or horizontal branch. The stars that  do not show any 
evidence of mixing   have  [C/N] $>$ $-0.60$ and lie on the lower RGB. 
 Figure 4 shows that all of our programme 
stars have [C/N] $>$ $-0.6$. This implies that none of our programme 
stars have undergone any significant internal mixing processes. 

 \begin{figure}[htb]
\centering
\includegraphics[width=8cm,height=8cm]{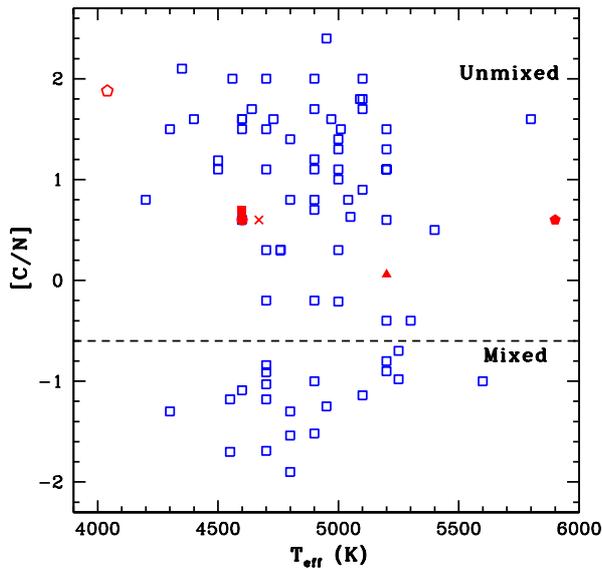}
\caption{Position of the programme stars in the [C/N] vs. T$_{eff}$ diagram. Programme stars are represented using red coloured symbols. The symbols used for the programme stars are same as in Figure 3. Open squares represent the stars from literature (Spite {\em et al.} 2006, Aoki {\em et al.} 2007, Goswami {\em et al.} 2016, Hansen {\em et al.} 2016b and Hansen {\em et al.} 2019). }
\end{figure}

\par Stellar models predict that when a low mass star ascends the red giant 
branch, the outer convective envelope expands inwards and penetrates into 
the region of CN processed materials (First dredge up (FDU)). The luminosity 
at which the FDU occurs in low mass field stars is
log(L/L$_{\odot}$) $\sim$ 0.80 (Gratton {\em et al.} 2000). All of our 
programme stars have log(L/L$_{\odot}$) $>$ 0.80, except  HD 176021. 
Hence they might have undergone first dredge up. However, these stars 
fall in the unmixed category. Even though in solar mass/metallicity 
stars, $^{12}$C/$^{13}$C ratio decreases by a factor of 20-30 from 
the original value and surface abundance of nitrogen increases after 
FDU (Iben \& Renzim 1984), it is found to be less efficient in 
metal-poor stars (VandenBerg \& Smith 1988, Charbonnel 1994). This implies 
that the changes in the surface compositions of C and N after the 
occurance of FDU are very small in metal-poor stars. The carbon abundance 
is found to be decreased only by about 0.05 dex and the decrease in 
the $^{12}$C/$^{13}$C ratio is not large enough for these stars 
(Gratton {\em et al.} 2000).   

\par A second mixing episode can happen when the star becomes brighter than 
the RGB bump at a luminosity of log(L/L$_{\odot}$) $\sim$ 2.0. In this 
process, $^{12}$C abundance decreases by a factor of $\sim$ 2.5 and 
$^{12}$C/$^{13}$C reaches a value of $\sim$ 6 to 10. Nitrogen abundance 
also increases by a factor of nearly 4 (Gratton {\em et al.} 2000). Two of 
our programme stars HD 30443 and CD$-38$ 2151 have 
log(L/L$_{\odot}$) $\sim$ 2.2 and 2.87 respectively. Hence they are 
expected to show the signatures of mixing. But these stars show no 
signatures of mixing based on [C/N] ratio. Spite {\em et al.} (2006) suggest 
that [C/N] raio is not a clean indicator of mixing as the abundance of 
C and N in the interstellar medium from which these stars are formed 
show large variations. In that case, one can use the $^{12}$C/$^{13}$C 
as a good indicator of mixing, since it is high in primordial matter 
($>$ 70) and the determination of carbon isotopic ratio is found to be 
insenstive to the choice of atmospheric parameters for the stars 
(Spite {\em et al.} 2006). We have plotted $^{12}$C/$^{13}$C ratio 
and T$_{eff}$ of our programme stars as shown in Figure 5. From the figure, 
it is clear that the object HD 30443 share the region occupied by the 
stars that have undergone internal mixing processes. This means that 
HD 30443 had experienced internal mixing processes that might have 
altered its initial surface chemical compositions. CD$-38$ 2151 lies 
close to the boundary of seperation of mixed and unmixed stars. The 
low value of carbon isotopic ratio in other programme stars may be 
interpretted as the result of strong processing in the AGB progenitors 
(Hansen {\em et al.} 2016b) and not in the star itself.  

 \begin{figure}[htb]
\centering
\includegraphics[width=8cm,height=8cm]{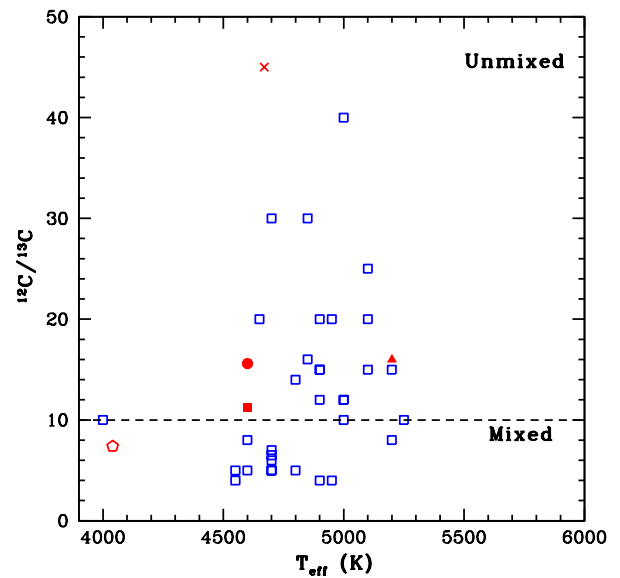}
\caption{Position of the programme stars in $^{12}$C/$^{13}$C vs. T$_{eff}$ diagram. Programme stars are represented using red coloured symbols. The symbols used for the programme stars are same as in Figure 3. Open squares represent the stars from Spite {\em et al.} (2006) and Aoki {\em et al.} (2007). }
\end{figure}
 
\par From the estimated carbon isotopic ratio and the location of our 
programme stars in the A(C) vs. [Fe/H] diagram, we assume that the 
observed enhancement 
of neutron-capture elements may be attributed to an extrinsic source. As 
Group I objects are mostly found to be associated with binary systems, the 
enhancement may be justified by the mass transfer from the binary companion. 

According to Busso {\em et al.}(2001), the light 
s-process elements such as Y, Zr and Sr are predominantly produced in AGB 
stars with near solar metallicity. Hence, mass transfer from such a companion 
can produce more enhancement of light s-process elements than the heavy 
s-process elements. The AGB stars with [Fe/H] $\sim$ $-1.0$ can produce 
more heavy s-process elements such as Ba, La, Ce and Nd than the light s-process 
elements, and this leads to  lower [ls/Fe] than [hs/Fe] in the stars that have 
accreted materials from the metal-poor AGB stars. The reason is that, in 
the metal-poor AGB stars, the number of Fe seed nuclei available for 
neutron-capture process is less and hence the neutron exposure for each 
Fe seed nuclei will be more. This leads to the formation of more heavier 
elements. The binary mass transfer scenario can similarly be applied to 
justify the observed enhancement of neutron-capture elements in 
CD$-28$ 1082 which is a CEMP-r/s star. Several authors 
(Hampel {\em et al.} 2016; Hansen {\em et al.} 2016a and references therein) 
have shown that the observed abundance patterns of CEMP-r/s stars are 
consistent with the model calculations of i-process in low-mettalicity AGB stars. 

\section{Conclusion}
The possible source of the origin of enhancement of neutron-capture elements 
in six carbon stars are examined in the light of absolute carbon abundances. Location 
of our programme stars in the A(C) vs. [Fe/H] diagram shows that these objects 
belong to Group I category. Various studies on radial vecities of these
objects show that most of them exhibit radial velocity variations. 
That is, these objects are mostly found to be associated with  binary systems. 
Hence the observed enhancement of carbon and neutron-capture 
elements in our programme stars may be attributed to the binary companion. 
But none of the programme stars are known to be confirmed binaries. 
The low values of carbon isotopic ratio also supports the extrinsic origin.  

\par Elemental abundance ratios also bear important signatures about the
source of enrichment of neutron-capture elements. We have re-estimated the
[hs/ls] ratio for our programme stars. Abate et al. (2015) predict an [ls/hs]
ratio less than zero for AGB models. All the programme stars except CD$-38$ 2151 and 
HD 176021 show an [hs/ls] ratio characteristics of AGB progenitors. 

\par We have also examined whether the programme stars have undergone
any internal mixing based on [C/N] and carbon isotopic ratios. Because
it is important to understand whether the mixing have modified the surface composition of
the star before interpretting the observed abundance patterns. 
We found that none of the proramme stars have experienced internal mixing, except HD 30443.  
The estimated values of [C/N] ratio also show that the objects have not gone through any 
significant mixing processes. Thus, the observed values of low carbon isotopic ratio
 in the unmixed stars may be due to the strong mixing processes occured in the AGB progenitors. 
In other words, the observed surface chemical compositions of our programme 
stars except HD 30443, preserve the fossil records of the materials 
synthesised in the AGB stars from which they have accreted the materials. 

\section*{Acknowledgements}
We thank the staff members at IAO, CREST and VBO for their assistance and 
cooperation during the observations. Funding from DST SERB project 
No. EMR/2016/005283 is gratefully acknowledged. This work made use of 
the SIMBAD astronomical database, operated at CDS, Strasbourg, France,  
the NASA ADS, USA and data from the European Space Agency (ESA) mission 
Gaia (https://www.cosmos.esa.int/gaia), processes by the Gaia Data Processing 
and Analysis Consortium 
(DPAC, https://www.cosmos.esa.int/web/gaia/dpac/consortium).


\begin{theunbibliography}{} 
\vspace{-1.5em}
\bibitem{latexcompanion}
Abate C., Pols O.R., Izzard R.G., Karakas A.I., 2015, A\&A, 581, A22
\bibitem{latexcompanion} 
Aoki W., Beers T.C., Christlieb N., Norris J.E., Ryan S.G et al., 2007, ApJ, 655, 492 
\bibitem{latexcompanion}
Asplund M., Grevesse N., Sauval A.J., 2009, Ann. Rev. Astron. Astrophy., 47:481
\bibitem{latexcompanion}
Bartkevicious A., 1996, Baltic Astron, 5, 217
\bibitem{latexcompanion}
Beers T.C., Preston G.W., Shectman S.A., 1985, AJ, 90, 2089
\bibitem{latexcompanion}
Beers T.C., Christlieb N., 2005, ARA\&A, 43, 531
\bibitem{latexcompanion}
Brooke J.S., Bernath P.F., Schmidt T.W., Bacskay G.B., 2013, J.Quant. Spectrosc. Radiat. Transfer, 124, 11
\bibitem{latexcompanion}
Busso M., Gallino R., Lambert D.L., Travaglio C., Smith V.V., 2001, ApJ, 557, 802
\bibitem{latexcompanion}
Charbonnel C., 2005, ASP Conference Series, Vol. 336
\bibitem{latexcompanion} 
Christlieb N., Green P. J., Wisotzki L., Reimers D., 2001, A\&A, 375, 366
\bibitem{latexcompanion}
Dearborn D. S. P., Lattanzio J. C., Eggleton P. P., 2006, ApJ, 639, 405
\bibitem{latexcompanion}
Eggleton P. P., Dearborn D. S. P., Lattanzio J. C., 2006, Science, 314, 1580
\bibitem{latexcompanion}
Frebel A., Aoki W., Christlieb N., Ando H., Asplund M. et al., 2005, Nature, 434, 871
\bibitem{latexcompanion}
Goswami A., 2005, MNRAS, 359, 531
\bibitem{latexcompanion}
Goswami A., Karinkuzhi D., Shantikumar N.S., 2010, MNRAS, 402, 1111
\bibitem{latexcompanion} 
Goswami A., Aoki W., Karinkuzhi D., 2016, MNRAS, 455, 402
\bibitem{latexcompanion}
Gratton R., Sneden C., Carretta E., Bragaglia A., 2000, A\&A, 354, 169
\bibitem{latexcompanion}
Hampel M., Stancliffe R.J., Lugaro M., Meyer B.S., 2016, ApJ, 831, 171 
\bibitem{latexcompanion}
Hansen C.J., Nordstr{\"o}m B., Hansen T.T., Kennedy C.R., Placco V M. et al., 2016b, A\&A, 588, A37
\bibitem{latexcompanion}
Hansen T. T., Andersen J., Nordstr{\"o}m B., Beers T.C., Placco V.M. et al. 2016a, A\&A, 586, A160
\bibitem{latexcompanion}
Hansen C.J., Hansen T.T., Koch A., Beers T.C., Nordstr{\"o}m B. et al., 2019, A\&A, 623, 128
\bibitem{latexcompanion}
Iben I.Jr., Renzini A., 1984, Phys. Letters 105, 329
\bibitem{latexcompanion}
Jonsell K., Barklem P.S., Gustafsson B., Christlieb N., Hill V. et al., 2006, A\&A, 451, 651
\bibitem{latexcompanion}
McClure R.D., 1983, ApJ, 208, 264
\bibitem{latexcompanion}
McClure R.D., 1984, ApJ, 280, 31
\bibitem{latexcompanion} 
McClure R.D., Woodsworth W., 1990, ApJ, 352, 709
\bibitem{latexcompanion}
McWilliiam A., 1998, AJ, 115, 1640
\bibitem{latexcompanion}
Norris J.E., Christlieb N., Korn A.J., Eriksson K., Bessell M.S. et al., 2007, ApJ, 670, 774
\bibitem{latexcompanion}
Placco V.M., Frebel A., Beers T. C., Stancliffe R. J., 2014, ApJ, 797, 21
\bibitem{latexcompanion}
Prochaska J.X., McWilliam A. 2000, ApJ, 537, 57
\bibitem{latexcompanion}
Purandardas M., Goswami A., Goswami P.P., Shejeelammal J., Masseron T., MNRAS, 2019, 486,3266
\bibitem{latexcompanion}
Purandardas M., Goswami A., Doddamani V.H., 2019, BSRSL, 88, 207
\bibitem{latexcompanion}
Ram R.S., Brooke James S.A., Bernath P.F., Sneden C., Lucatello S., 2014, ApJS, 211, 5
\bibitem{latexcompanion} 
Sneden C., 1973, PhD thesis, Univ. Texas
\bibitem{latexcompanion} 
Sneden C., Lucatello S., Ram R.S., Brook J.S.A., Bernath P., 2014, Ap. J. Supp., 214, 26
\bibitem{latexcompanion}
Spite M., Cayrel R., Plez B., Hill V., Spite F. et al., 2005, A\&A 430, 655
\bibitem{latexcompanion}
Spite M., Cayrel R., Hill V, Spite F., Francois P. et al., 2006, A\&A 455, 291
\bibitem{latexcompanion}
Spite, M., Caffau, E., Bonifacio, P., Spite F., Ludwig H.-G. et al. 2013, A\&A, 552, 107
\bibitem{latexcompanion}
VandenBerg D.A., Smith G.H., 1988, PASP 100, 314
\bibitem{latexcompanion}
Yong D., Norris J. E., Bessell  M. S., Christlieb N., Asplund M. et al., 2013, ApJ, 762, 26
\bibitem{latexcompanion}
Yoon J., Beers T.C., Placco V.M., Rasmussen K.C., Carollo D. et al., 2016, ApJ, 833, 20
\bibitem{latexcompanion}
Vanture A.D., 1992, AJ, 104, 1977
\bibitem{latexcompanion}
Wisotzki L., Christlieb N., Bade N., Beckmann V., K{\"o}hler T, et al., 2000, A\&A, 358, 77
\bibitem{latexcompanion} 
Worley C.C., Hill V.J., Sobeck J., Carretta E., 2013, A\&A, 553, A47
\end{theunbibliography}
\end{document}